\documentclass[review]{elsarticle}

\usepackage{lineno,hyperref}
\usepackage{amssymb,amsmath}
\usepackage{paralist}
\usepackage{booktabs}
\allowdisplaybreaks[1]
\modulolinenumbers[5]

\journal{Commun. Nonlinear Sci. Numer. Simulat.}

\usepackage{numcompress}

\begin{document}

\begin{frontmatter}

\title{Coupling-induced oscillations in two intrinsically quiescent populations}

\author{Almaz Mustafin\fnref{myfootnote}}
\address{Kazakh National Technical University, 22 Satpayev St., Almaty 050013, Kazakhstan}
\fntext[]{E-mail address: mustafin\_a1@kazntu.kz; mustafin@mailaps.org; Tel.: +7(727)227~61~21}




\begin{abstract}
A model of two consumer-resource systems linked by interspecific interference competition of consumers is considered. The basic assumption of the model is that the dynamics of the resource is much slower than that of the consumer. In the absence of interaction each consumer-resource pair has a unique stable steady state which is completely nonoscillatory. When weakly coupled, the consumer-resource pairs are shown to exhibit sustained low-frequency synchronous antiphase relaxation oscillations.
\end{abstract}


\end{frontmatter}


\section{Introduction}
Recently, synchronization of coupled oscillators has been a subject of extensive study, not only due to the ubiquity of this phenomenon, but also owing to importance of its applications in engineering and biology \cite{Hoppensteadt:1997,Pikovsky:2001,Strogatz:2003,Vandermeer:2006,Balanov:2009}. The conventional assumption of the theory of synchronization is that in the uncoupled state each elementary unit of the linked system is oscillatory. However no less interesting are the systems where coupling is essential for the very emergence of oscillations and not only for their synchronization and phase adjustment.

The first example of coupling-induced periodicity has been propounded by Smale \cite{Smale:1974}. His abstract model of a biological cell involves chemical kinetics of four metabolites, such that the reaction equations for the set of metabolites have a globally stable equilibrium. The cell is ``dead'', in that the concentrations of its metabolites always relax to the same fixed levels. When two such cells are coupled by linear diffusion terms, however, the resulting equations are shown to have a globally stable limit cycle. The concentrations of the metabolites begin to oscillate, and the system becomes ``alive''.

Since that time, triggered by Smale's seminal work, a number of plausible models have been proposed in which coupling of identical nonoscillating cells of concrete nature could generate synchronous oscillations. The majority of these models concern neural cells with excitable membrane \cite{Loewenstein:2001,Gomez-Marin:2007,Szatmari:2008}. Szatm{\'{a}}ri and Chua \cite{Szatmari:2008} suggested an apt term ``awakening dynamics'' for the phenomenon.

The subject of the present paper is an emergence of collective oscillations in a simple system of two coupled nonoscillatory consumer-resource pairs. Our choice of coupled consumer-resource equations as a matter of enquiry is dictated primarily by the abundance and importance of consumer-resource relations. Consumer-resource communities are the building bricks of ecosystems. Depending on a specific nature of the involved consumer-resource interactions, they can take the forms of predator-prey, herbivore-plant, parasite-host, and exploiter-victim systems \cite{Murdoch:2003}. However applications of the consumer-resource models extend far beyond the ecology and are found wherever one can speak of win-loss interactions. In its broad meaning, resource is any substance which can lead to increased growth rate of the consumer as its availability in the environment is increased. As this takes place, the resource is certainly consumed. Consuming the resource means tending to reduce its availability. When carefully examined, consumer-resource models are identified in the following fields: epidemiology (infected and susceptible \cite[ch.~10]{Murray:2002}), laser dynamics (photons and electrons \cite[ch.~6]{Carroll:1985}), labor economics (share of labor and employment rate \cite[p.~28]{Zhang:1991}), theoretical immunology (antigens and B lymphocytes \cite[p.~299]{Volkenstein:1983}), kinetics of chain chemical reactions (free radicals and lipid molecules \cite{Chernavskii:1977}), and in numerous other studies from diverse disciplines.

So far as we know, examples of coupling-induced synchronization of intrinsically nonoscillatory populations have never been proposed. We are going to show that interaction in a form of density-dependent cross-losses may drive two nonoscillatory consumer-resource pairs into synchronous periodic pulsing.

\section{The model}
Of all types of interactions between individuals of the same population (intraspecific interactions) or individuals of different populations (interspecific interactions) of the same trophic level competition is most commonly encountered. In a broad sense, competition takes place when each species (individual) has an inhibiting effect on the growth of the other species (individual). An inhibiting effect should be understood to mean either an increase in the death rate or a decrease in the birth rate.

Consider the famous consumer-resource equations proposed by MacArthur \cite{MacArthur:1970,Chesson:1990}:
\begin{subequations}\label{CRMac}
\begin{alignat}{3}
\dot{x}_{j}& = \bigl[r_{j}(1 -{x_{j}}/{K_{j}}) -\sum\nolimits_{i=1}^{n}c_{ij}y_{i}\bigr]x_{j},\quad & j&=1,\dotsc,m,\label{CRMac-x}\\
\dot{y}_{i}& = \Bigl(\sum\nolimits_{j=1}^{m}c_{ij}w_{j}x_{j} -b_{i}\Bigr)y_{i},\quad & i&=1,\dotsc,n.\label{CRMac-y}
\end{alignat}
\end{subequations}
Here dots indicate differentiation with respect to time $t$, $x_{j}$ represents the total biomass of $j$th resource (prey), $y_{i}$ stands for the total biomass of $i$th consumer (predator) species, the constant $r_{j}$ defines the growth rate of $j$th resource, $K_{j}$ is the carrying capacity of $j$th resource, $c_{ij}$ is the rate of uptake of a unit of $j$th resource by each individual of $i$th consumer population, $w_{j}^{-1}$ is the conversion efficiency parameter representing an amount of $j$th resource an individual of $i$th consumer population must consume in order to produce a single new individual of that species, $b_{i}$ is the loss rate of $i$th consumer due to either natural death or emigration. All parameters in \eqref{CRMac} are nonnegative.

MacArthur \cite{MacArthur:1970} assumed population dynamics of the resources to be much faster than that of the consumers which enabled him to approximate $x_{j}$ in \eqref{CRMac-y} by its quasi-steady-state value derived by setting the right-hand side of \eqref{CRMac-x} to zero. As a result, he succeeded in reducing slow-scale subsystem of equations \eqref{CRMac-y} to the well-known Lotka--Volterra--Gause (LVG) model \cite{Gause:1935}
\begin{equation}\label{VG}
\dot{y}_{i} = \Bigl(k_{i} -\sum\nolimits_{s=1}^{n}a_{is}y_{s}\Bigr)y_{i},\quad i =1,\dotsc,n,
\end{equation}
where $a_{is}= \sum\nolimits_{j=1}^{m}c_{ij}c_{sj}(w_{j}K_{j}/r_{j})$ and $k_{i}= \sum\nolimits_{j=1}^{m}c_{ij}w_{j}K_{j} -b_{i}$ $(i,s=1,\dotsc,n)$. Resources do not enter LVG equations explicitly being parameterized by carrying capacities.

More recently, such an asymptotic reduction has also been carried out for a model of competition where species (with continuous trait) consume the common resource that is constantly supplied, under the assumption of a very fast dynamics for the supply of the resource and a fast dynamics for death and uptake rates \cite{Mirrahimi:2013}.

Consumer-resource model \eqref{CRMac} assumes that competition between consumer species is purely exploitative: individuals and populations interact through utilizing (or occupying) a common resource that is in short supply. Quite on the contrary, LVG model \eqref{VG} describes competition strictly phenomenologically, as direct interference where consumers experience harm attributed to their mutual presence in a habitat (e.g. through aggressive behavior). However we should stress that neither does MacArthur's reduction claim that interference competition entirely results from ``more fundamental'' trophic competition, nor does it urge us to hastily consider direct competition as some ``derived'' concept. What it states is that when the dynamics of the consumers are associated with a slow time scale,  the effects of exploitation competition are indistinguishable from those of interference competition. And at slow-time scale, coefficients $a_{is}$ of \eqref{VG} merely would add to ``true'' interference coefficients $a'_{is}$ if the interference is accounted for properly in \eqref{CRMac-y}.

Most mathematical models dealing with coupled consumer-resource pairs or multilevel trophic chains ignore contributions of intraspecific and interspecific interference effects in consumers. Indeed, the empirical data like \cite{Devetter:2008} do indicate that $a'_{ij}$ may be negligible in comparison with $a_{is}$. Nevertheless, literature advocating the explicit accounting for direct interference shows that incorporation of self-limitation and cross-limitation terms in the equations at the consumers' level can provide for the stable coexistence of many species on few resources [\citealp{Kirlinger:1986}; \citealp[p.~31]{Bazykin:1998}; \citealp{Kuang:2003}].

Moreover, if we are to assume dynamics of the resources to be much slower than that of the consumers, it is likely that we have to introduce interference competition terms in subsystem \eqref{CRMac-y}.

Consider the following modification of \eqref{CRMac} representing coupled two-consumer, two-resource equations:
\begin{subequations}\label{CRcoupled}
\begin{align}
\dot{x}_{1}& = p_{1} -(c_{1} y_{1} +q_{1}) x_{1},\label{CRcoupled-x1}\\
\dot{x}_{2}& = p_{2} -(c_{2} y_{2} +q_{2}) x_{2},\label{CRcoupled-x2}\\
\dot{y}_{1}& = (c_{1} w_{1} x_{1} -b_{1} -d_{1} y_{1} -h_{2}y_{2})y_{1},\label{CRcoupled-y1}\\
\dot{y}_{2}& = (c_{2} w_{2} x_{2} -b_{2} -d_{2} y_{2} -h_{1}y_{1})y_{2}.\label{CRcoupled-y2}
\end{align}
\end{subequations}

Instead of the logistic mode of resource supply, as is the case in MacArthur's model, our model is based on so-called ``equable'' mode of resource exploitation \cite{StewartLevin:1973}, by which the quantities of available resources are held constant by a continuous-flow system. According to  \eqref{CRcoupled-x1} and \eqref{CRcoupled-x2}, a constant concentration of $j$th resource ($j=1,2$) flows into a defined volume with the rate $p_{j}$ while unused resource flows out with the specific rate $q_{j}$, in much the same manner as in a chemostat \cite{Herbert:1956}. In natural conditions, the equable modes of feeding, for instance, can be found on the first trophic level of ecosystem, among autotrophs.

We assume that the consumers' functional response is linear. Endowing the consumers with nonlinear (and even distinct from one another) functional responses seems a premature complication of the model. Such an extension of MacArthur's generic model, as shown by Abrams and Holt \cite{Abrams:2002}, may lead to several notable modes of behavior, including coexistence via periodic cycling. However, the resulting coupled oscillations in their model are rather entrained than ``awakened'', because one of the two involved species, when unlinked, is able to oscillate.

Besides, in \eqref{CRcoupled-y1} and \eqref{CRcoupled-y2} intraspecific competition strength $d_{i}$ ($i=1,2$) measures direct interference of individuals within $i$th consumer population with each other resulting in an additional per capita loss rate $d_{i} y_{i}$; interspecific competition strength $h_{s}$ ($s=1,2$; $s\neq i$) quantifies direct interference effect from $s$th consumer on $i$th consumer resulting in an additional per capita loss rate, $h_{s}y_{s}$, of the latter.

Equations \eqref{CRcoupled} contain two important assumptions. First, they assume that the resources are noninteractive. On higher trophic levels, however, resources may interact and the possibility of competition among the resources was originally pointed out by Levine\cite{Levine:1976} and empirically confirmed by Lynch \cite{Lynch:1978}. Since then, a whole series of theoretical papers (based on MacArthur's equations) have been published on two-predator, two-prey systems with interference competition between two self-reproducting prey species \cite{Vandermeer:1980,Kirlinger:1986,Xiang:2006}.

As seen from \eqref{CRcoupled-x1} and \eqref{CRcoupled-x2}, there is no intraspecific interference competition within the resource populations, in distinction to MacArthur's model. Yet the resource abundance would remain finite even in the absence of the consumer.

The second assumption of our equations is that the consumers interact only directly, through interference competition. They cannot compete trophically, through their use of resources, as each consumer specializes on one resource only. In the models of pure trophic competition, like MacArthur's logistic-supply model \cite{Hsu:1979} and equable-supply model \cite{StewartLevin:1973,Tilman:1982}, each predator is allowed to feed on both prey.

Intraspecific interference competition is allowed within the consumers as well. Owing to this assumption, a consumer would remain bounded even though the abundance of the associated resource happened to be constant.

The novelty of model \eqref{CRcoupled} is that it considers time hierarchy of MacArthur's consumer-resource equations to be reversed by assuming dynamics of the consumers to be much faster than that of the involved resources and articulates the importance of direct competition mechanisms within the framework of this assumption.

\begin{figure}
\noindent\centering{
\includegraphics[scale=0.5]{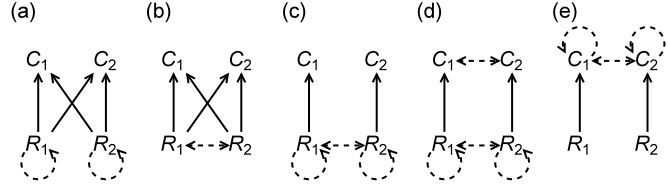}}
\caption{Variations on a theme of MacArtur's model as proposed by different authors. $C_{1}$ and $C_{2}$ are consumers, $R_{1}$ and $R_{2}$ are resources. Solid arrows indicate predation, dashed mean interference. (a) \cite{MacArthur:1970,Hsu:1979}; (b) \cite{Levine:1976,Vandermeer:1980}; (c) \cite{Kirlinger:1986,Xiang:2006}; (d) \cite{Kirlinger:1986}; (e) present paper.}
\label{fig1}
\end{figure}

The distinguishing features of aforementioned versions of generic MacArthur's model, including the system under consideration, are summarized diagrammatically on Fig.~\ref{fig1}.

Upon the scaling $u_{1} = {c_{1}w_{1}x_{1}}/{b_{1}} -1$, $u_{2} = {c_{2}w_{2}x_{2}}/{b_{2}} -1$, $v_{1} = {c_{1}y_{1}}/{q_{1}}$, $v_{2} = {c_{2}y_{2}}/{q_{2}}$, $\gamma_{1} = {c_{1}p_{1}w_{1}}/{b_{1}q_{1}} -1$, $\gamma_{2} = {c_{2}p_{2}w_{2}}/{b_{2}q_{2}} -1$, $\delta_{1} = {d_{1}q_{1}}/{b_{1}c_{1}}$, $\delta_{2} = {d_{2}q_{2}}/{b_{2}c_{2}}$, $\varkappa_{1} = {h_{1}q_{1}}/{b_{2}c_{1}}$, $\varkappa_{2} = {h_{2}q_{2}}/{b_{1}c_{2}}$, $\beta = {q_{1}}/{q_{2}}$, $\varepsilon_{1} = {q_{1}}/{b_{1}}$, $\varepsilon_{2} = {q_{2}}/{b_{2}}$, and $t' = q_{1}t$, equations \eqref{CRcoupled} take the following nondimensional form:
\begin{equation}\label{CRnondim}
\begin{split}
\dot{u}_{1}& =\gamma_{1} -u_{1}v_{1} -u_{1} -v_{1},\\
\beta\dot{u}_{2}& =\gamma_{2} -u_{2}v_{2} -u_{2} -v_{2},\\
\varepsilon_{1}\dot{v}_{1}& =(u_{1} -\delta_{1}v_{1} -\varkappa_{2}v_{2})v_{1},\\
\varepsilon_{2}\dot{v}_{2}& =(u_{2} -\delta_{2}v_{2} -\varkappa_{1}v_{1})v_{2}.
\end{split}
\end{equation}
Note that in \eqref{CRnondim} dots mean differentiation with respect to nondimensional ``slow'' timescale variable $t'$ measured in units of the resource lifetime $1/q_{1}$, as defined by the chosen scaling.

The parameters $\beta^{-1}$, $\varepsilon_{1}^{-1}$ and $\varepsilon_{2}^{-1}$ reflect the rapidity of the dynamics of $u_{2}$, $v_{1}$ and $v_{2}$ with reference to that of $u_{1}$. It is assumed that $\beta=\mathcal{O}(1)$, $\varepsilon_{1},\varepsilon_{2}\ll 1$ and $\delta_{1},\delta_{2} \ll 1$.

For the sake of simplicity but without any loss of generality, we set $\beta=1$, $\varepsilon_{1}=\varepsilon_{2}=\varepsilon$ and $\delta_{1}=\delta_{2}=\delta$, and also drop the prime at $t$. At the same time, we retain resource income rates, $\gamma_{1}$ and $\gamma_{2}$, and coupling strengths, $\varkappa_{1}$ and $\varkappa_{2}$, as free control parameters of the model. Eventually model equations take the form
\begin{equation}\label{coupled-uv}
\begin{split}
\dot{u}_{1}& = \gamma_{1} -(u_{1} +1)v_{1} -u_{1},\\
\dot{u}_{2}& = \gamma_{2} -(u_{2} +1)v_{2} -u_{2},\\
\varepsilon \dot{v}_{1}& = (u_{1} -\delta v_{1} -\varkappa_{2}v_{2})v_{1},\\
\varepsilon \dot{v}_{2}& = (u_{2} -\delta v_{2} -\varkappa_{1}v_{1})v_{2}.
\end{split}
\end{equation}

It should be mentioned that being proportional to its dimensional prototype, $v_{i}$ directly represents population density of consumer species and is always nonnegative. Quantity $u_{i}$, however, is not an abundance of resource in the true sense of the word. It is rather an affine transformation of $x_{i}$ done for reasons of mathematical convenience. Unlike a purely linear transformation, an affine map does not preserve the zero point, so in \eqref{coupled-uv} $u_{i} = -1$ corresponds to zero level of $i$th resource in reality. Nevertheless, from here on we will apply the term ``resource'' to $u_{i}$ for brevity.

\section{Analysis and implications}
When $\varkappa_{1},\varkappa_{2} = 0$, the consumers are independent. An uncoupled consumer-resource system obeys the equations
\begin{equation}\label{isolated-uv}
\begin{split}
\dot{u}& = \gamma -(u +1)v -u,\\
\varepsilon \dot{v}& = (u -\delta v)v,
\end{split}
\end{equation}
which have a unique stable positive steady state:
\begin{equation}\label{steady-uv}
\begin{split}
\overline{u}& = \textstyle{\frac{1}{2}}[\sqrt{1 +(4\gamma +2 +\delta)\delta} -1 -\delta] =\gamma\delta +\mathcal{O}(\delta^{2}),\\
\overline{v}& = \textstyle{\frac{1}{2\delta}}[\sqrt{1 +(4\gamma +2 +\delta)\delta} -1 -\delta] =\gamma -\gamma(\gamma +1)\delta +\mathcal{O}(\delta^{2}).
\end{split}
\end{equation}
Equilibrium \eqref{steady-uv} is a node (intrinsically nonoscillatory steady state) for $\varepsilon = o(\delta^{2})$. In the subsequent discussion we assume that this condition is fulfilled.

Physically feasible equilibria, $(\overline{u}_{1},\overline{u}_{2},\overline{v}_{1},\overline{v}_{2})$, of \eqref{coupled-uv} are those for which $\overline{v}_{1},\overline{v}_{2}\geqslant 0$. We denote the interior fixed point by $F_{12}=(\overline{u}_{1}, \overline{u}_{2},\overline{v}_{1},\overline{v}_{2})$, where the subscripts at ``$F$'' stand for the consumers. Lack of a certain index at a boundary fixed point means that the consumer concerned is not present (extinct). Thus $F_{1}=(\overline{u}_{1},\overline{u}_{2},\overline{v}_{1},0)$ and $F_{2}=(\overline{u}_{1},\overline{u}_{2},0,\overline{v}_{2})$ designate either of one-consumer equilibria corresponding to dominance, while $F=(\overline{u}_{1},\overline{u}_{2},0,0)$ means both consumers having been washed out.

Model \eqref{coupled-uv} has four feasible steady states. To $\mathcal{O}(1)$ for small $\delta$
\begin{equation}\label{ss}
\begin{split}
F:\quad&\overline{u}_{1}=\gamma_{1},\quad \overline{u}_{2}=\gamma_{2},\quad \overline{v}_{1}=0,\quad \overline{v}_{2}=0;\\
F_{1}:\quad&\overline{u}_{1}=0,\quad \overline{u}_{2}=\gamma_{2},\quad \overline{v}_{1}=\gamma_{1},\quad \overline{v}_{2}=0;\\
F_{2}:\quad&\overline{u}_{1}=\gamma_{1},\quad \overline{u}_{2}=0,\quad \overline{v}_{1}=0,\quad \overline{v}_{2}=\gamma_{2};\\
F_{12}:\quad&\overline{u}_{1}=\frac{\varkappa_{1}\gamma_{1} -\varkappa_{2}\gamma_{2} -\varkappa_{1}\varkappa_{2} +1 \pm R}{2(\varkappa_{1} -1)},\\
&\overline{u}_{2}=\frac{-\varkappa_{1}\gamma_{1} +\varkappa_{2}\gamma_{2} -\varkappa_{1}\varkappa_{2} +1 \pm R}{2(\varkappa_{2} -1)},\\
&\overline{v}_{1}={\overline{u}_{2}}/{\varkappa_{1}},\quad \overline{v}_{2}={\overline{u}_{1}}/{\varkappa_{2}},\\
\text{where}\quad&R=\bigl[(\varkappa_{1}\gamma_{1} -\varkappa_{2}\gamma_{2} -\varkappa_{1}\varkappa_{2} +1)^{2}\bigr.\\
& \phantom{R=\bigl[}\bigl.+4\varkappa_{2}(\varkappa_{1} -1)(\varkappa_{1}\gamma_{1} -\gamma_{2})\bigr]^{1/2}.
\end{split}
\end{equation}

\begin{table}
\caption{Existence and stability conditions of nonnegative equilibria in system \eqref{coupled-uv}.}
\begin{center}
\begin{tabular}{lll}
\toprule
Equilibrium & Existence & Stability \\
\midrule
$F$       & Always & Never \\
$F_{1}$   & Always & $\gamma_{2}/\gamma_{1}<\varkappa_{1}$ \\
$F_{2}$   & Always & $\gamma_{2}/\gamma_{1}>1/\varkappa_{2}$ \\
$F_{12}$  & $1/\varkappa_{2}<\gamma_{2}/\gamma_{1}<\varkappa_{1}$&\\
          & for\ $\varkappa_{1},\varkappa_{2}>1$\ (strong coupling) &Never\\
          & $\varkappa_{1}<\gamma_{2}/\gamma_{1}<1/\varkappa_{2}$&\\
          & for\ $\varkappa_{1},\varkappa_{2}<1$ (weak coupling) &$\varkappa_{1},\varkappa_{2}=o(\varepsilon^{1/2})$\\
\bottomrule
\end{tabular}
\end{center}
\label{tab1}
\end{table}
Existence and stability conditions of equilibria \eqref{ss} are summarized in Table \ref{tab1}. The model reveals qualitatively different behavior at strong and weak coupling between consumer-resource pairs. For lack of space we are not able to discuss the case of strong coupling in detail and restrict ourselves to a brief comment. If coupling is strong, any static coexistence of competing consumers is not possible: one of the consumers wins and completely dominates. Intense competition makes possible bistability of boundary equilibria, as evident from Table \ref{tab1}. When both $F_{1}$ and $F_{2}$ are stable with an unstable coexistence steady state $F_{12}$, the system being studied is able to exhibit a hysteresis effect. In the case of bistability the winner is determined by the initial conditions.

Application of Routh--Hurwitz stability criterion yields that $F_{12}$ is stable for fairly small coupling strengths, $\varkappa_{1},\varkappa_{2}=o(\varepsilon^{1/2})$. However, $\varepsilon$ is so small, that from the practical viewpoint, $F_{12}$ turns out to be unstable for any moderately weak (physically reasonable) coupling. As seen from Table \ref{tab1}, the very existence of the interior equilibrium $F_{12}$ in the case of weak coupling, $\varkappa_{1},\varkappa_{2}<1$, implies instability of both boundary fixed points, $F_{1}$ and $F_{2}$. System \eqref{coupled-uv} happens to possess four nonnegative steady states, none of them being stable. The instability of $F_{12}$ is through growing oscillations. In such a case, the model would thus be expected to have a limit cycle in its four-dimensional phase space corresponding to sustained oscillations. For sufficiently weak coupling strengths of order $\mathcal{O}(\varepsilon^{1/2})$, i.\,e. not too far away from the Hopf bifurcation, this limit cycle is small and represents a low-amplitude quasi-harmonic periodic solution. As a practical matter, the range of such an infinitesimally weak coupling is of less concern to us than is the range of far more feasible not-too-weak coupling, corresponding to well-developed substantially nonlinear oscillations.

By the assumption, $0<\varepsilon \ll 1$, meaning that system \eqref{coupled-uv} is singularly perturbed. The slow variables are resources, $u_{1}$ and $u_{2}$, and the fast variables are consumers, $v_{1}$ and $v_{2}$. The standard practice of reducing such systems is multiple-scale analysis \cite{Verhulst:2005} whereby fast variables are adiabatically eliminated. One has to establish the validity of the adiabatic elimination in each specific case. In particular, Tikhonov's theorem requires quasi-steady states of the fast equations to be stable.

To decompose the full system \eqref{coupled-uv} into fast and slow subsystems, introduce fast time variable $\tau=t/\varepsilon$. Now rescale \eqref{coupled-uv} by replacing $t$ with $\tau\varepsilon$ and, after taking $\varepsilon=0$, it becomes
\begin{equation}\label{fast}
\begin{split}
u'_{1}& = u'_{2} = 0,\\
v'_{1}& = (u_{1} -\delta v_{1} -\varkappa_{2}v_{2})v_{1},\\
v'_{2}& = (u_{2} -\delta v_{2} -\varkappa_{1}v_{1})v_{2},
\end{split}
\end{equation}
where prime means differentiation with respect to $\tau$. This is the fast subsystem, where $u_{1}$ and $u_{2}$ are replaced by their initial values and treated as parameters. It yields the inner solution, valid for $t = \mathcal{O}(\varepsilon)$.

Setting $\varepsilon=0$ in \eqref{coupled-uv} leads to the slow subsystem
\begin{subequations}\label{slow}
\begin{align}
\dot{u}_{1}& = \gamma_{1} -(u_{1} +1)v_{1} -u_{1},\label{slow-u1}\\
\dot{u}_{2}& = \gamma_{2} -(u_{2} +1)v_{2} -u_{2},\label{slow-u2}\\
0& = (u_{1} -\delta v_{1} -\varkappa_{2}v_{2})v_{1},\label{slow-v1}\\
0& = (u_{2} -\delta v_{2} -\varkappa_{1}v_{1})v_{2},\label{slow-v2}
\end{align}
\end{subequations}
which produces the outer solution, valid for $t = \mathcal{O}(1)$. In this singular limit as $\varepsilon \to 0$, the subsystem defines a slow flow on the surface (slow manifold) given by \eqref{slow-v1} and \eqref{slow-v2}. Outer solution is valid for those $u_{1}$ and $u_{2}$, for which the quasi-steady states of the fast subsystem \eqref{fast} are stable.

We anticipate the dynamics of the full system \eqref{coupled-uv} in its four-dimensional phase space $(u_{1},u_{2},v_{1},v_{2})$ to consist of two typical motions: quickly approaching the slow manifold \eqref{slow-v1} and \eqref{slow-v2}, and slowly sliding over it until a leave point (where the solution disappears) is reached. After that, the representing point may possibly jump to another local solution of \eqref{slow-v1} and \eqref{slow-v2}.

Thus, we ought to find all quasi-steady states of the fast subsystem \eqref{fast}, map the domains of their stability onto the slow phase plane $(u_{1},u_{2})$, and then investigate the dynamics of the slow subsystem \eqref{slow} with piecewise continuous functions.

The fast subsystem \eqref{fast}, which is nothing but the classical LVG model, has four quasi-steady states---three boundary and one interior---denoted by $Q$ (the slow variables are deemed to be frozen):
\begin{subequations}\label{qss}
\begin{alignat}{3}
Q:\quad &\widetilde{v}_{1}=0,\quad && \widetilde{v}_{2}=0;\label{qssQ}\\
Q_{1}:\quad &\widetilde{v}_{1}={u_{1}}/{\delta},\quad && \widetilde{v}_{2}=0;\label{qssQ1}\\
Q_{2}:\quad &\widetilde{v}_{1}=0,\quad && \widetilde{v}_{2}={u_{2}}/{\delta};\label{qssQ2}\\
Q_{12}:\quad &\widetilde{v}_{1}=\frac{\varkappa_{2}u_{2} -\delta u_{1}}{\varkappa_{1}\varkappa_{2} -\delta^{2}},
\quad && \widetilde{v}_{2}=\frac{\varkappa_{1}u_{1} -\delta u_{2}}{\varkappa_{1}\varkappa_{2} -\delta^{2}}.\label{qssQ12}
\end{alignat}
\end{subequations}
Quasi-equilibria $Q_{1}$ and $Q_{2}$ are stable nodes respectively for $\delta u_{2} < \varkappa_{1}u_{1}$ and $\delta u_{1} < \varkappa_{2}u_{2}$; $Q$ and $Q_{12}$ are always unstable.

\begin{figure}
\noindent\centering{
\includegraphics[width=2.5in]{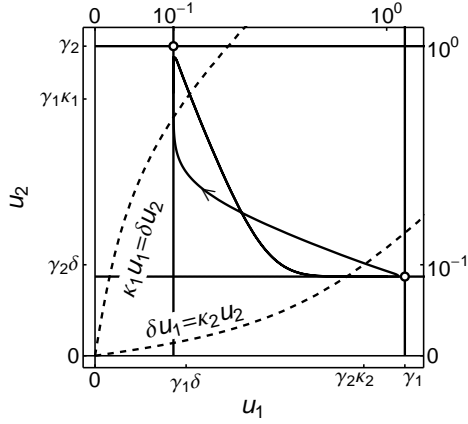}}
\caption{The limit cycle of system \eqref{coupled-uv} projected onto phase plane of the slow variables (resources $u_{1}$ and $u_{2}$). The direction of motion is indicated by an arrow. Consumer 1 is in a stable nonzero quasi-equilibrium for any combination of resources below the line $\varkappa_{1} u_{1} -\delta u_{2} = 0$, consumer 2---above the line $\delta u_{1} -\varkappa_{2}u_{2} = 0$. For better appearance we perform scaling $u_{i} \to \operatorname{arsinh}(u_{i}/\gamma_{i}\delta)$ ($i=1,2$).}
\label{fig2}
\end{figure}
Consider the plane of resources shown on Fig.~\ref{fig2}. Let initially $Q_{1}$ be stable and $Q_{2}$ unstable with consumer 1 dominating. This corresponds to slow variables $u_{1}$ and $u_{2}$ being somewhere below the line $\delta u_{1} -\varkappa_{2}u_{2} = 0$. The dynamics of the resources (treated as bifurcation parameters in reference to the consumers) is described by a system of two independent equations
\begin{equation}\label{slow-piecewise1}
\begin{split}
\dot{u}_{1}& = \gamma_{1} -\bigl[(u_{1} +1)/\delta +1\bigr]u_{1},\\
\dot{u}_{2}& = \gamma_{2} -u_{2},
\end{split}
\end{equation}
which, in view of \eqref{qssQ1}, is a piecewise version of the slow subsystem \eqref{slow}. System \eqref{slow-piecewise1} has a stable steady state
\begin{equation}\label{ss-u1u2-1}
\begin{split}
\widehat{u}_{1}^{(1)}&=\textstyle\frac{1}{2}[\sqrt{1 +(4\gamma_{1} +2 +\delta)\delta} -1 -\delta] = \gamma_{1}\delta +\mathcal{O}(\delta^{2}),\\
\widehat{u}_{2}^{(1)}&=\gamma_{2}.
\end{split}
\end{equation}
It is marked by an open circle in the upper left corner of Fig.~\ref{fig2}. While heading to \eqref{ss-u1u2-1}, the trajectory crosses the line $\delta u_{1} -\varkappa_{2}u_{2} = 0$ and enters the domain of bistability of both $Q_{1}$ and $Q_{2}$. However the dominance of consumer 1 persists.

Note that in \eqref{slow-piecewise1}, the variable $u_{1}$ is faster than $u_{2}$ due to small $\delta$. Clearly, the representing point must have relaxed to the vertical line $u_{1}=\widehat{u}_{1}^{(1)}\approx \gamma_{1}\delta$ well before approaching the horizontal line $u_{2}=\widehat{u}_{2}^{(1)}=\gamma_{2}$.

Eventually the trajectory has to cross the line $\varkappa_{1}u_{1} -\delta u_{2} =0$. As soon as this has happened, node $Q_{1}$ of the fast subsystem \eqref{fast} will be absorbed by saddle $Q_{12}$. Consumer 1 rapidly washes out, and $Q_{2}$ becomes the only stable quasi-equilibrium, with consumer 2 dominating.

In terms of the four-dimensional phase space of full system \eqref{coupled-uv}, the representing point is now in the other stable branch of the slow manifold given by \eqref{slow-v1} and \eqref{slow-v2}. The motion over this alternative branch obeys the piecewise subsystem
\begin{equation}\label{slow-piecewise2}
\begin{split}
\dot{u}_{1}& =\gamma_{1} -u_{1},\\
\dot{u}_{2}& =\gamma_{2} -\bigl[(u_{2} +1)/\delta +1\bigr]u_{2}
\end{split}
\end{equation}
with the initial conditions $u_{1}(0)=\gamma_{1}\delta$ and $u_{2}(0)=\gamma_{1}\varkappa_{1}$. The dynamics of \eqref{slow-piecewise2} is basically similar to that of \eqref{slow-piecewise1} analyzed above. The variable $u_{2}$ comparatively rapidly relaxes to $\widehat{u}_{2}^{(2)} = \gamma_{2}\delta +\mathcal{O}(\delta^{2})$; the variable $u_{1}$ slowly grows toward $\widehat{u}_{1}^{(2)} = \gamma_{1}$. When $u_{1}$ has crossed the level $\gamma_{2} \varkappa_{2}$, node $Q_{2}$ would be absorbed by saddle $Q_{12}$. The system returns to the first branch of the slow manifold, and thereby the oscillatory cycle gets closed.

\section{Results and discussion}
\begin{figure}
\noindent\centering{
\includegraphics[width=4in]{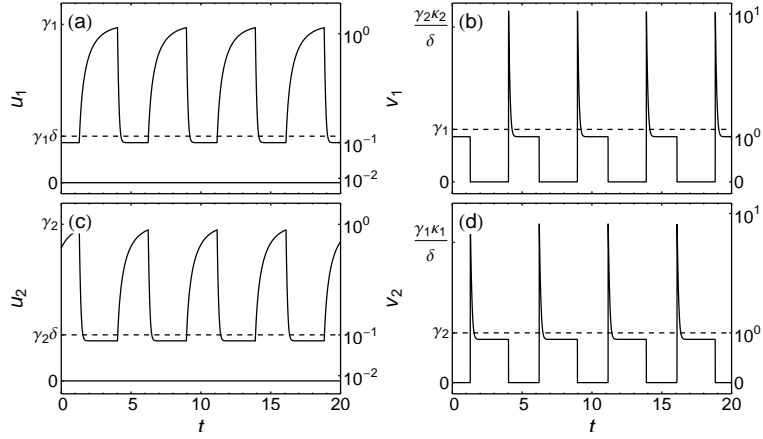}}
\caption{Time profiles of self-sustained relaxation oscillations experienced by coupled consumer-resource pairs. (a) resource 1, (b) consumer 1, (c) resource 2, and (d) consumer 2. The numerical values of the parameters are $\varepsilon = 0.73{\times}10^{-3}$, $\gamma_{1} = 1.2$, $\gamma_{2} = 1$, $\varkappa_{1}=0.5$, $\varkappa_{2}=0.8$, and $\delta = 0.1$.}
\label{fig3}
\end{figure}

As seen from Fig.~\ref{fig2}, the two coupled consumer-resource communities execute self-sustained synchronous antiphase-locked oscillations. Fig.~\ref{fig3} shows the results of numerical integration of \eqref{coupled-uv}. The resources demonstrate sawtooth periodic pulses. The oscillation range for the resource levels remains finite and, what is important, it does not depend on the intraspecific interference parameter $\delta$. The consumers change periodically between extinction\footnote{Actually, the exact solution to full system \eqref{coupled-uv} yields nonzero $v_{1}$ and $v_{2}$ at any time, even though they may take on very small values, so that $\ln (v) \propto -\varepsilon^{-1}$. It should be emphasized that the competing consumers periodically ``die out'' only within the framework of a multiple-scale technique's approximation being used. This is merely convenient idealization.} and respective constant levels $\gamma_{1}$ and $\gamma_{2}$. Very brief transient from zero to flat nonzero level within each cycle is accompanied by a highly pronounced spiky overshoot. The magnitude of the spike tends to infinity as $\delta \to 0$, in view of \eqref{qssQ1} and \eqref{qssQ2}.

One may distinguish four parts within the period of synchronous oscillations:
\begin{compactenum}[1)]
\item Consumer 1 is essentially zero, while consumer 2 is approximately equal to its uncoupled steady-state value, $\gamma_{2}$. Resource 1 increases due to its constant inflow until it overcomes losses for consumer 1;
\item With a sufficient resource stock, consumer 1 now emerges. The population 1 exhibits a spike due to the fast time scale of the consumer equations. The sharp increase in population saturates the available resource level, so resource 1 drops. Cross-losses cause consumer 2 to wash out;
\item Quantities $v_{1}$ and $u_{1}$ relax to their equilibrium values, as if there were only one uncoupled consumer-resource pair. Consumer 2 is essentially zero. Resource 2 is increasing, like resource 1 did in part 1;
\item Resource 2 surpasses the losses, consumer 2 emerges and the subsequent cross-losses cause consumer 1 to wash out. The spiking consumer 2 also causes a substantial decrease in the available stock of the associated resource. The sequence begins again.
\end{compactenum}
The essential feature of the model is that when one consumer is very scarce, the whole coupled system behaves like an uncoupled consumer-resource pair \eqref{isolated-uv}.

Presenting his famous model Smale remarked that ``it is more difficult to reduce the number of chemicals to two or even three'' \cite[p. 26]{Smale:1974}. As distinct from Smale's example, the bilinear coupling in our case makes self-sustained synchronous oscillations possible for just two variables.

As we have seen, phase trajectory of the system constantly moves from the neighborhood of unstable boundary equilibrium $F_{1}$ where only consumer 1 is present, to the neighborhood of $F_{2}$ where consumer 2 completely dominates, back to $F_{1}$, and so on in cyclic alternation. This kind of trajectory was termed ``heteroclinic cycle'' by Kirlinger \cite{Kirlinger:1986}. A heteroclinic cycle occurs when the outflow (unstable manifold) from one saddle point is directly connected to the inflow (stable manifold) of another saddle point, and vice versa. It is closely related to another notion of the nonlinear dynamics, a homoclinic cycle, which emerges when the unstable and the stable manifolds of the same saddle coincide and form a closed loop.

\begin{figure}
\noindent\centering{
\includegraphics[scale=1]{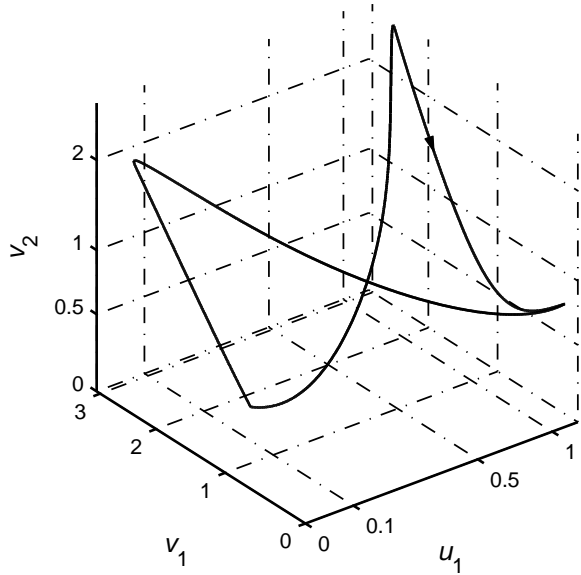}}
\caption{A 3D-projection of the limit cycle in system \eqref{coupled-uv} for parameters chosen in a neighborhood of the heteroclinic cycle.}
\label{fig4}
\end{figure}

Homo- and heteroclinic cycles are not robust structures in the sense that infinitesimally small change of system parameters destroy them. However in the practical sense, any limit cycle passing in close proximity to saddle points will be indistinguishable from a heteroclinic cycle (Fig.~\ref{fig4}). The only difference is strict periodicity, although the period of the limit cycle in a neighborhood of the heteroclinic cycle may be long. Besides, at the threshold of homo-/heteroclinic bifurcation the period is susceptible to external noise.

In the context of our model, as coupling becomes stronger, the stable limit cycle swells and passes closer and closer to boundary fixed points which are node-saddles. Depending on the interplay between the parameters, eventually it may bang into one or both of these equilibria creating either a homoclinic or heteroclinic cycle, respectively. This corresponds to $\gamma_{2}/\gamma_{1}=\varkappa_{1}$ and $\gamma_{1}/\gamma_{2}=\varkappa_{2}$. On further increasing the coupling, the saddle connection breaks and the loop is destroyed.

It is worth noting that heteroclinic cycles were first found by May and Leonard \cite{May:1975} in a classical LVG system with competing three species. However in their model the cycle is not truly periodic: as time goes on, the system tends to stay in the neighborhood of any one boundary equilibrium ever longer, so that the ``total time spent in completing one cycle is likewise proportional to the length of time the system has been running.'' Moreover, May and Leonard state that ``the phenomenon clearly requires at least three competitors, which is why it cannot occur in models with two competitors.'' This statement is echoed by Vandermeer \cite{Vandermeer:2011} who extended their theory on higher dimensions: ``It appears to be the case that all cases of an odd number of species follow this basic pattern, whereas all cases of even number of species result in extinction of half of the components, leaving the other half living independently at their carrying capacities.'' In view of our results, the above conclusion is by far and away true providing one stays within the framework of classical LVG equations, which in fact implies a high rapidity of the resource dynamics. In our model of just two competitors the slowness of the resource relative to the consumer is essential for the oscillations to occur, because it provides the necessary inertia to the system.

The feasibility of our model is tightly bound to justification of the adopted time hierarchy in system \eqref{coupled-uv}. In ecosystems, the most common case is rapid consumption of food by species. However it seems reasonable to propose that the model may describe the first level of an ecosystem, at which the consumers are autotrophs and the resources are mineral nutrients. The ability to exploit different substrates leads to a possibility of stable coexistence of different organisms descending from a common ancestor. Divergent evolution is just the emergence of new species: due to mutations two populations come into being, sharing the same genetic code but having proteins able to process different substrates. Providing the environmental conditions are quite stable on the evolutionary timescale, the inflows of inorganic substrates from the surroundings may be considered constant and the washout time of a substrate may occur much longer than the life expectancy of a species (recall the definition $\varepsilon=q/b$).

From a non-ecological perspective, by and large similar relationships can be found in coupled longitudinal modes of laser with second harmonic generation. Baer \cite{Baer:1986} experimentally observed antiphase oscillations of two (and more) modes in a multimode neodymium-doped yttrium aluminum garnet ($\mathrm{Nd^{3+}{:}Y_{3}Al_{15}O_{12}}$) laser with an intracavity potassium titanyl phosphate ($\mathrm{KTiOPO_{4}}$) frequency-doubling crystal. He also proposed a model for that phenomenon in terms of the coupled rate equations. In original designations, the equations for the population inversions, $G_{1}$ and $G_{2}$, and the intensities, $I_{1}$ and $I_{2}$, are as follows:
\begin{subequations}\label{Baer}
\begin{align}
\tau_{f}\dot{G}_{1}& = G_{1}^{0} -(\beta_{1}I_{1} +\beta_{12}I_{2} +1)G_{1},\label{Baer-G1}\\
\tau_{f}\dot{G}_{2}& = G_{2}^{0} -(\beta_{2}I_{2} +\beta_{21}I_{1} +1)G_{2},\label{Baer-G2}\\
\tau_{c}\dot{I}_{1}& = (G_{1} -\alpha_{1} -\varepsilon I_{1} -2\varepsilon I_{2})I_{1},\label{Baer-I1}\\
\tau_{c}\dot{I}_{2}& = (G_{2} -\alpha_{2} -\varepsilon I_{2} -2\varepsilon I_{1})I_{2},\label{Baer-I2}
\end{align}
\end{subequations}
where $\tau_{f}$ and $\tau_{c}$ are the fluorescence time and cavity round trip time, respectively; $\beta_{1}$ and $\beta_{2}$ are self-saturation parameters, which determine how strongly the corresponding intensity depletes the available gain; $\beta_{12}$ and $\beta_{21}$ are cross-saturation parameters; $G_{1}^{0}$ and $G_{2}^{0}$ are the small signal gains (pump parameters); $\alpha_{1}$ and $\alpha_{2}$ are the cavity losses for the respective modes; and $\varepsilon$ is the nonlinear coupling coefficient due to the presence of the intracavity doubling crystal. In \eqref{Baer-I1} and \eqref{Baer-I2}, the second-order terms $I_{1}^{2}$, $I_{2}^{2}$  and $I_{1}I_{2}$ account for the loss in intensity of the fundamental frequencies through second harmonic generation and sum-frequency generation.

It is apparent at a glance that \eqref{Baer} represents coupled consumer-resource equations, where the intensities play the role of consumers, while the population inversions act as resources. Structurally, Baer's system is very nearly the same as system \eqref{CRcoupled} being discussed. Even the time hierarchy is similar: $\tau_{c}/\tau_{f}=(0.5\ \mathrm{ns})/(0.24\ \mathrm{ms})=\mathcal{O}(10^{-6})$, i.\,e. the ``resources'' change much slower than the ``consumers''. There are three distinctions, however. First, in \eqref{Baer-I1} and \eqref{Baer-I2}, intra- and interspecific interference parameters are not independent, both of them being proportional to the coupling strength. Second, in \eqref{Baer-G1} and \eqref{Baer-G2}, the two modes are allowed to ``compete'' for the active medium, so the consumer-resource pairs turn out to be further linked trophically. Third, when uncoupled, the steady state of each mode is stable focus, not stable node as is the case in our model. In other words, uncoupled consumer-resource pairs in Baer's system are weakly damped oscillators with the intrinsic period proportional to $\sqrt{\tau_{c}\tau_{f}}$.

Baer performed numerical integration of \eqref{Baer} for different coupling strength $\varepsilon$ and with many different initial conditions. He revealed that as $\varepsilon$ decreases, the mode-coupling oscillation period decreases tending to the period of intrinsic oscillations. If $\varepsilon$ is decreased further, the oscillations cease, and the system becomes stable. With a large $\varepsilon$, the oscillation period becomes quite long, and each mode appears to reach a stable intensity value before abruptly switching off (cf. the results of our analysis!). The numerical solutions correctly predicted that the two modes tend to pulse on and off out of phase with each other.

Subsequently, Erneaux and Glorieux \cite[pp.~318--325]{Erneux:2010} reduced \eqref{Baer} to the equations for coupled quasi-conservative oscillators and proved the existence of stable antiphase periodic solution in the case of the modes with identical parameters $G^{0}$, $\alpha$ and $\beta$. However that result has to do with the onset of low-amplitude quasi-harmonic oscillations. Unlike their study, our approach deals with well-developed high-magnitude essentially nonlinear oscillations.

Interesting issues concern how the outcome of exploitation and interference is altered when the mode of resource supply is not constant, or when the interspecific interference is not necessarily mutually costly (i.\,e. each consumer suffers a net reduction in per capita growth rate via interference from, but can gain an increase in growth rate via interference on, the other consumer) \cite{Amarasekare:2003}. Investigations of these possibilities may constitute a future direction for work on the model.

\section{Conclusions}
We considered a model of two consumer-resource pairs linked by interspecific interference competition. When uncoupled, an individual consumer-resource pair has a unique stable steady state and does not admit periodic solutions. If intraspecific interference within the species is strong enough, the equilibrium is nonoscillatory.

When coupling is moderately weak, the model reveals low-frequency antiphase relaxation oscillations. The consumers cannot coexist even dynamically: in each of two periodically alternating states one consumer completely dominates and the other is on the verge of extinction. The most intriguing feature of the model is that each of the involved consumer-resource pairs taken separately does not oscillate; both communities are completely quiescent, however, in interaction, when coupled in a nonlinear way, the resulting system turns into a relaxation oscillator.

\section*{Acknowledgement}
The author wishes to express his gratitude to the referees for their valuable suggestions.

\section*{References}

\end{document}